# Effective acetylene length dependence of the elastic properties of different kinds of graphynes

*Guilherme B. Kanegae, Alexandre F. Fonseca* *

Applied Physics Department, Institute of Physics "Gleb Wataghin", University of Campinas - UNICAMP, 13083-859, Campinas, São Paulo, Brazil.

**Corresponding Author**

* Phone: +55 19 3521-5364. Email: Alexandre F. Fonseca – afonseca@ifi.unicamp.br

**ABSTRACT:** Graphyne is a planar network of connected carbon chains, each formed by $n$ acetylene linkages. Uncountable ways to make these connections lead to uncountable structural graphyne families (GFs). As the synthesis of graphynes with $n > 1$ has been reported in literature, it is of interest to find out how their physical properties depend on $n$ for each possible GF. Although literature already present specific models to describe the dependence on $n$ of the elastic properties of specific GFs, there is not yet enough amount of data for the physical properties of different graphynes with different values of $n$. Based on fully atomistic molecular dynamics simulations, the Young's modulus, shear modulus, linear compressibility and Poisson's ratio of 10 graphyne members of 7 different GFs are calculated. A simple elastic model consisting of a serial combination of $n$ springs is proposed to describe the dependence on $n$ of the elastic properties of these 7 GFs. We show that except for the Poisson's ratio, this simple unique elastic model is able to numerically describe, with good precision, the Young's modulus, shear modulus and linear compressibility of all different graphynes, including anisotropy and negative values of linear compressibility of some GFs.

# 1. Introduction

Graphyne (GY) is a planar carbon allotrope whose structure can be thought as a network of acetylene chains of length $n$, $(-[-C\equiv C-]_n-)$, directly connected to themselves, by aromatic rings, and/or sp2 carbon-carbon bonds [1-5]. Despite originally proposed in the 80's [1], the interest on GYs grew in the last decade. GYs have been shown to possess non-null band-gap [1-8] and good [1,6-13] or even better [8,14,15] electronic properties than graphene. In-plane stiffness [16-23] and thermal conductivity [10,24-27] of GYs were shown to be smaller than that of graphene. These interesting physical properties of GYs have motivated the development of possible applications in thermoelectric devices [24,25] and nanotransistors [28]. In addition, the porosity of GYs makes them suitable for applications in anodes of batteries [29], desalination [30,31], hydrogen storage [32], among others [2-4,33].

Structures known as graph*di*ynes, graph*tri*ynes, etc., correspond to GYs with $n$ = 2, 3, etc., respectively. Different physical properties of GYs having $n \geq 2$ have been reported in the literature [2-5,22,24,34-38]. In particular, studies about the dependence on $n$ of the mechanical properties of the so-called "γ-" [39,40], "β-" [41] and "α-" [42] GYs have been recently published. The synthesis of GYs with $n$ = 2 [43-47] and $n$ = 4 [48,49] have been also reported in the literature. These theoretical and experimental efforts reveal the interest towards the development of a more general comprehension of the physical properties of GYs. In this work, we present computational results for the dependence on $n$ of the Young's modulus, shear modulus, Poisson's ratio and linear compressibility of 10 GY members ($1 \leq n \leq 10$) of the 7 different families of GYs (or here called "GFs"), as originally proposed by Baughman *et al*. [1], including anisotropic GYs, totaling data for 70 GY structures. This data is, then, used to test available models for the dependence on *n* of the elastic properties of specific GFs [39-42]. After that, we demonstrate the possibility to use a more general and simple *effective elastic model* to numerically describe (with good precision) the elastic moduli and linear compressibility of all kinds of GYs, including the anisotropic ones. Only the dependence on *n* of the Poisson's ratio of all GFs is shown to require more specific models. Such an *effective elastic model* would be of practical interest not only by providing a simple qualitative mechanism behind the elastic behavior of GY structures, but also by allowing more easily predictions of these properties in possible applications. Counterintuitively, it will be shown that a model as simple as a serial combination of *n* equal springs is more effective to describe the dependence on *n* of the elastic properties of different GYs than a more sophisticated model that includes the hinging and bending of the GY acetylene chains. The effectiveness of this simple elastic model is demonstrated by comparing its ability to fit, with high precision, the dependence on *n* of the Young's modulus of all GYs with that of the more sophisticated model. Also, the *effective elastic model* is shown to provide with good precision, and without any additional fitting, the shear moduli and the linear compressibility of all GYs. Although the dependence on *n* of the Poisson's ratio of GY structures cannot be described by this simple model, it turned out to be surprising that notwithstanding the dependence of the linear compressibility on the Poisson's ratio, the dependence on *n* of the linear compressibility of GYs is shown to be precisely described by the simple *effective elastic model* of a serial combination of *n* springs. In addition, our results predict negative linear compressibility of two families of GYs along the zigzag direction (only one work in the literature predicted negative linear compressibility for one particular family of GYs [50], and only for structures with $n \leq 5$), and the value of 0.95 for the Poisson's ratio of the α-graphyne with $n = 10$, that is amongst the largest values ever predicted for 2D symmetric crystals [51].

To make clear the presentation of our results, we adopted a notation similar to that of Ivanovskii [2] who numbered the GFs from 1 to 7, instead of using prefixes like "γ-", "6,6,12-", "β-", or "α-" GYs. Here, each simulated GY structure will be named as "G$n$Y$f$", with $n$ and $f$ indicating the number of acetylene linkages in the chains, $1 \leq n \leq 10$, and the corresponding GF, $1 \leq f \leq 7$, according to Ivanovskii notation, respectively. Figure **1** displays the structures of the 7 GFs with $n = 1$, or the "G1Y$f$" structures.

This work is organized as follows. In the next section, we describe the theory used to obtain the elastic properties of all GYs and the computational methods to obtain them. In Section III, the results are presented and discussed, including the proposed analysis of the effectiveness of a simple elastic model for all GYs. In Section IV, we present the concluding remarks.

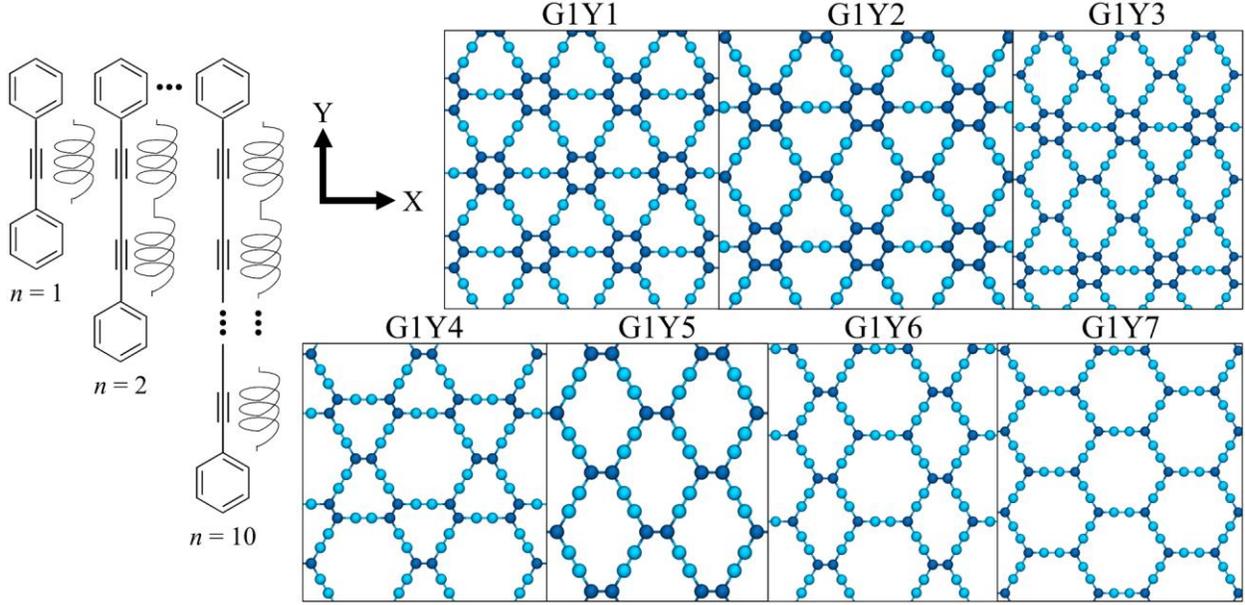

**Figure 1**: Left: illustration of the dependence on *n* of a piece of GY structure. Springs are drawn to illustrate the *effective elastic model* as described in the text. Right: G1Y*f* structures of the seven originally proposed GFs [1]. Horizontal, x, and vertical, y, axes correspond to "armchair" and "zigzag" directions, respectively. *f* goes from 1 to 7, and represents the same sequence given by Ivanovskii [2]. G1Y1, G1Y2, G1Y4 and G1Y7 are commonly called "γ-", "6,6,12-", "β-", and "α-" graphynes, respectively. The structures of the corresponding G2Y*f*s, G3Y*f*s, etc., are similar to those shown but with *n* = 2, 3, etc., respectively. Cyan (blue) circles represent carbon atoms of acetylene chains (sp2 carbon atoms).

## 2. Theory and computational methods

The in-plane elastic properties of GYs will be obtained following the method described by Andrew *et al.* [52]. Two-dimensional crystals admit four non-null elastic constants, $C_{11}$, $C_{12}$, $C_{22}$ and $C_{66}$ (in Voigt notation), whose elastic energy per unit area is given by:

$$U(\varepsilon) = \tfrac{1}{2}C_{11}\varepsilon_{xx}^2 + \tfrac{1}{2}C_{22}\varepsilon_{yy}^2 + C_{12}\varepsilon_{xx}\varepsilon_{yy} + 2C_{66}\varepsilon_{xy}^2 , \tag{1}$$

where $\varepsilon_{xx}$ and $\varepsilon_{yy}$ are strains along x and y directions, respectively, and $\varepsilon_{xy} = \frac{1}{2}\left(\frac{du_x}{dy} + \frac{du_y}{dx}\right)$ [53,54] is the shear strain where $u_x$ e $u_y$ are displacements along x and y directions, respectively. Once the $C_{ij}$ of all G$n$Y$f$s are obtained, their corresponding elastic properties are given by:

$$E_x = \frac{C_{11}C_{22} - C_{12}^2}{C_{22}}, \quad E_y = \frac{C_{11}C_{22} - C_{12}^2}{C_{11}}, \tag{2}$$

$$G = C_{66}, \quad v_{xy} = \frac{C_{12}}{C_{22}}, \quad v_{yx} = \frac{C_{12}}{C_{11}}, \tag{3}$$

$$\beta_x = \frac{C_{22} - C_{12}}{C_{11}C_{22} - C_{12}^2}, \quad \beta_y = \frac{C_{11} - C_{12}}{C_{11}C_{22} - C_{12}^2}, \tag{4}$$

where $E_x$ ($E_y$) and $\beta_x$ ($\beta_y$) are the Young's modulus and the linear compressibility along x (y) direction, respectively, $G$ is the shear modulus, and $v_{xy}$ ($v_{yx}$) is the Poisson's ratio of the structures when longitudinal strain is applied along x (y) direction and the resultant transversal strain is measured along y (x) direction. Similar to what was made in Ref. [22], the calculations of the $C_{ij}$ of the G$n$Y$f$ structures are not made previously assuming that they are or not symmetric, in order to verify the consistence of the results from the computational method with respect to what is expected regarding symmetry. G$n$Y1, G$n$Y4 and G$n$Y7 (G$n$Y2, G$n$Y3, G$n$Y5 and G$n$Y6) GFs are symmetric (asymmetric), and $E_x = E_y \equiv E$, $v_{xy} = v_{yx} \equiv v$ and $\beta_x = \beta_y \equiv \beta$ for symmetric structures.

To obtain the first three elastic constants, $C_{11}$, $C_{12}$ and $C_{22}$, three series of simulations are performed with each structure: two *uniaxial* and one *biaxial* tensile numerical experiments. The uniaxial tensile experiment consists of fixing the dimension along one direction, and obtaining the energy of the structure as a function of strain along the other direction. From eq. (1), we get $U_\text{x}^\text{uniaxial} \equiv U(\varepsilon_{xx} \equiv \varepsilon, \varepsilon_{yy} = 0) = 0.5 C_{11} \varepsilon^2$ ($U_\text{y}^\text{uniaxial} \equiv U(\varepsilon_{xx} \equiv 0, \varepsilon_{yy} = \varepsilon) = 0.5 C_{22} \varepsilon^2$) that allows us to obtain $C_{11}$ ($C_{22}$) by fitting a set of values of the energy $U$ versus $\varepsilon$, provided the maximum $\varepsilon$ is within the linear elastic regime of deformation of the structure. The biaxial experiment consists of applying the same amount of strain along x and y directions at the same time, i. e., $\varepsilon_{xx} = \varepsilon_{yy} \equiv \varepsilon$. From eq. (1), we get the equations $U^\text{biaxial} \equiv U(\varepsilon_{xx} = \varepsilon_{yy} = \varepsilon) = M\varepsilon^2$ and $C_{12} = M - 0.5(C_{11} + C_{22})$, needed to obtain $C_{12}$ once the values of $C_{11}$ and $C_{22}$ were already obtained. The elastic constant $C_{66}$ can be obtained by series of shear strain experiments as follows. With the dimensions of the structure along both x and y directions kept fixed, i.e. $\varepsilon_{xx} = \varepsilon_{yy} = 0$, the shear is applied on the structure along x direction by making $\frac{du_x}{dy} = \varepsilon$ and $\frac{du_y}{dx} = 0$. In this case, $\varepsilon_{xy} = \frac{1}{2}\left(\frac{du_x}{dy} + \frac{du_y}{dx}\right) = \frac{1}{2}\varepsilon$ and eq. (1) becomes $U^\text{shear} \equiv U(\varepsilon_{xx} = \varepsilon_{yy} = 0, \varepsilon_{xy} = 0.5\varepsilon) = 0.5 C_{66} \varepsilon^2$. In all simulations, the maximum $\varepsilon = 1\%$.

These tensile and shear computational experiments were performed through energy minimizations of the structure under periodic boundary conditions for applied strain increments of 0.1% consistent with the type of test as described above. The energy minimization is based on conjugate-gradient algorithm implemented in LAMMPS package [55], with energy and force tolerances set to $10^{-8}$ and $10^{-8}$ eV/Å, respectively. To ensure the convergence of the energy minimization procedure to the smallest energy structure, an additional protocol was developed based on the proposal given in Ref. [56]. According to Sihn *et al*. [56] in order to ensure the energy minimization algorithm in LAMMPS, it is important to add a dynamical test. The idea is to allow the structure to dynamically evolve without any constraint for few timesteps (we run about 1000 timesteps), after the application of the algorithm of energy minimization. Then, verify how much kinetic energy the system acquires. Smaller the kinetic energy of the structure after this short dynamic simulation closer the structure to that of smallest energy. The threshold value of $10^{-8}$ eV for this kinetic energy is chosen as a criterion of optimization. A loop of energy minimizations followed by short dynamical runs of the structure was, then, developed and applied to each structure at each value of strain.

The AIREBO potential [57,58] was employed to simulate the carbon-carbon interactions. It has been successfully used to study mechanical and thermal properties of GYs of $n = 1$ and $n = 2$ [18,22,35,59], with good agreement of the structural parameters with DFT calculations. The sizes of the systems varied from about 35 Å (for $n = 1$) to 208 Å (for $n = 10$) in order to have supercells formed by several unit cells of each structure.

## 3. Results and Discussion

The $C_{ij}$ obtained from the MD simulations and the elastic properties calculated by eqs. (2)-(4), of all G$n$Y$f$s are presented in Table **S1** of Supplementary Material (SM). Here, we will verify the agreement between our MD data and the available models in the literature for the dependence on $n$ of the elastic properties of "γ-" [39,40], "β-" [41] and "α-" [42] GYs. Then, adding fitting parameters to two models, one based on the spring model for the elastic properties of GYs as given by Ref. [39], and the other based on the more sophisticated model that includes stretching, hinging and bending of the acetylene chains, given by Ref. [42], we numerically investigate their effectiveness to describe the MD data for all GFs.

### 3.1 Test of the available elastic models for specific GYs

In this subsection, we will verify the relative precision of the available specific models for the dependence on $n$ of the Young's modulus, shear modulus and Poisson's ratio of "γ-" [39,40], "β-" [41] and "α-" [42] GYs. For each specific model, we will plot the corresponding

MD data together with the curve representing the model. For each curve, we calculate the coefficient of determination, $R^2$. While the model used by Cranford *et al*. [39] is based on a simple serial combination of *n* springs, the models given by Hou *et al*. [40,42] and Qu *et al*. [41] are given in terms of stretching, hinging and bending of the acetylene chains. Amongst these models, only the first does not provide expressions for the dependence on *n* of the shear modulus and Poisson's ratio of the corresponding GF.

### 3.1.1 The elastic models for the G*n*Y1 structures

Cranford *et al*. [39] and Hou *et al*. [40] proposed models for the dependence on *n* of the Young's modulus of "γ-"GYs (or our G*n*Y1 family). Hou *et al*. [40] provided also expressions for the shear modulus and Poisson's ratio of G*n*Y1 structures. For the Cranford *et al*. [39] model, we assumed the validity of an expression for the dependence on *n* of the shear modulus of G*n*Y1 structures similar to that of Young's modulus. Cranford *et al*. [39] model requires the determination of the lattice parameter of the G*n*Y1 structures. The lattice parameters of all G*n*Y*f* structures were calculated based on the MD results and are given in Section **S2** of the SM. Figure **2** shows the plots of their expressions and the MD data. The values of $R^2$ for the Cranford *et al*. (Hou *et al*.) model of G*n*Y1 structures are 0.9710 and 0.9574 (0.9132 and 0.6180), for the Young's and shear moduli, respectively. As the curves for the Poisson's ratio do not agree with the MD data, $R^2$ is not calculated for that. Cranford et al. [39] did not predicted the dependence on *n* of the Poisson's ratio of G*n*Y1 structures.

It is important to say that Cranford *et al*. [39] have not calculated the shear modulus of G*n*Y1 structures, so the results confirm that their model is able to capture the physics behind their elastic properties.

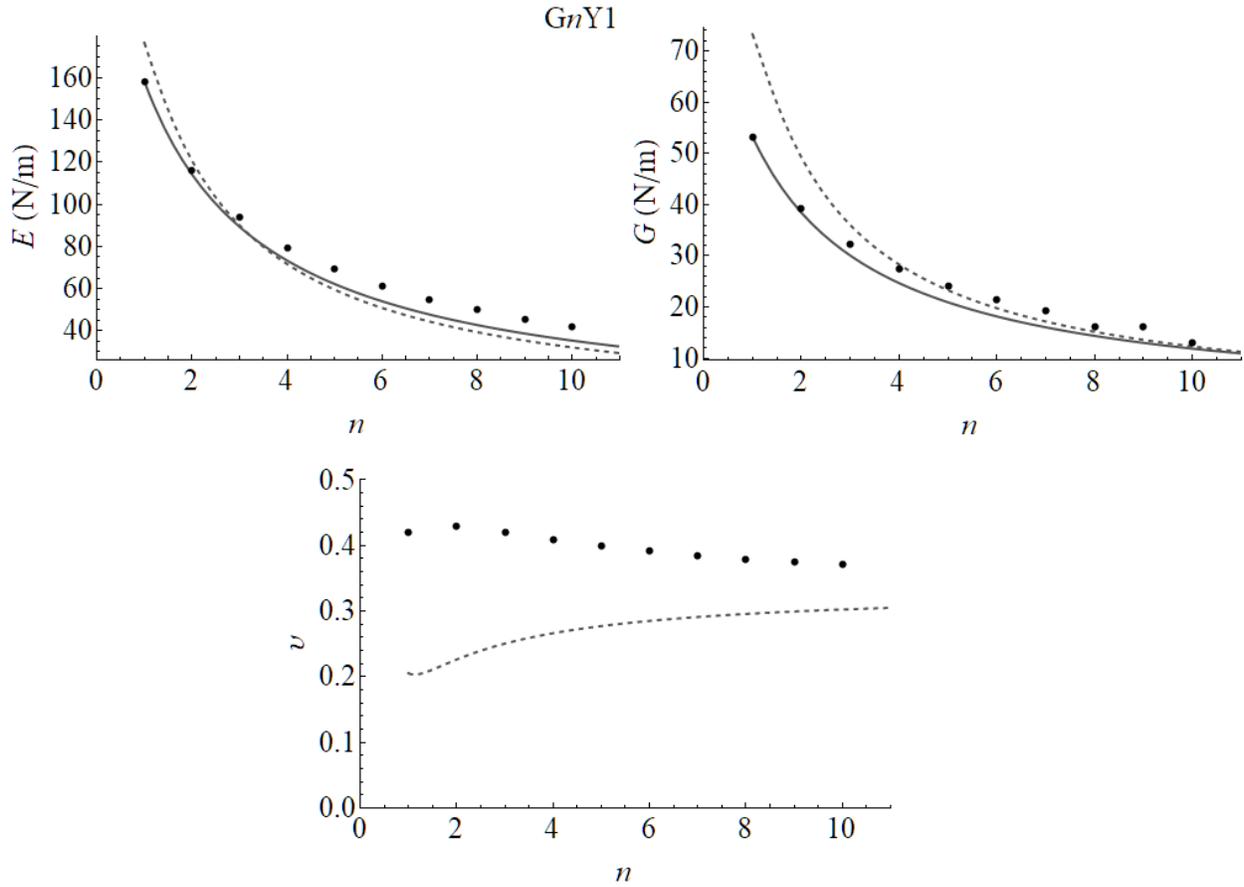

**Figure 2**: Young's modulus (top left), shear modulus (top right) and Poisson's ratio (bottom) of all "γ-" or G$n$Y1 structures as function of $n$. MD data is shown in circles. Full lines represent the curves given by eq. (9) of Cranford *et al*. [39] model that is valid for Young's or shear modulus, depending on $Y_1$ of their eq. (9) being the Young's or shear modulus of the G1Y1 structure. Dashed lines represent the eq. (15) (left panel), eq. (17) (right panel) and eq. (16) (bottom panel) of the Hou *et al*. [40] model.

### 3.1.2 The elastic model for G$n$Y4 structures

Qu *et al*. [41] proposed a model for the dependence on $n$ of the Young's modulus, shear modulus and Poisson's ratio of "β-"GYs (or our G$n$Y4 family). Figure **3** shows the curves and the MD data. Although the dependence on $n$ of the Young's modulus of G$n$Y4 structures are well described by Qu *et al.* [41] model with $R^2 = 0.9190$, shear modulus and Poisson's ratio are clearly not well described by their model ($R^2$ for the shear modulus, for example, is < 0).

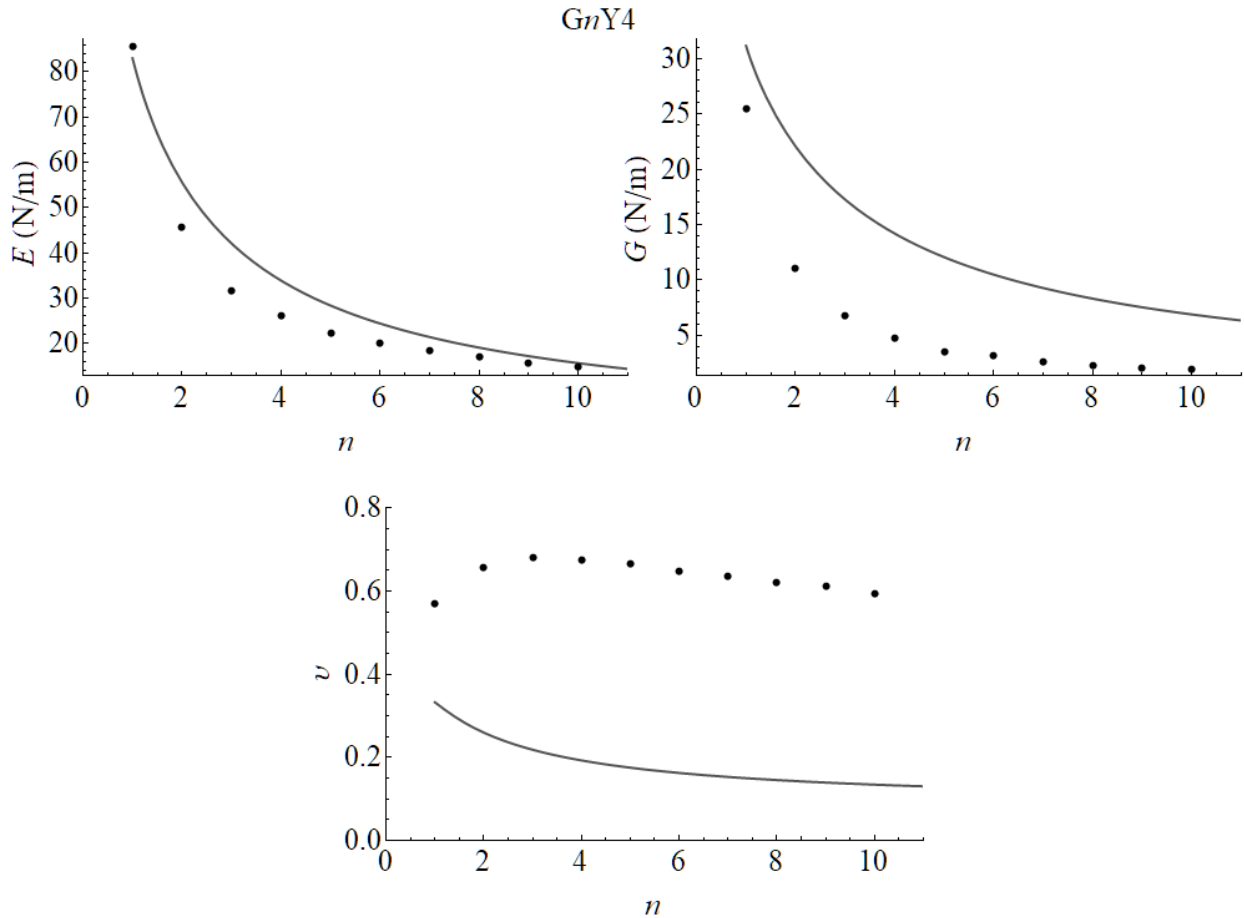

**Figure 3**: Young's modulus (top left), shear modulus (top right) and Poisson's ratio (bottom) of all "β-" or G$n$Y4 structures as function of $n$. MD data is shown in circles while full lines, from top left to bottom, represent the curves given by eq. (13), (15) and (14) of Qu *et al.* [41] model, respectively.

### 3.1.3 The elastic model for G$n$Y7 structures

Hou *et al.* [42] proposed a model for the dependence on $n$ of the Young's modulus, shear modulus and Poisson's ratio of "α-"GYs (or our G$n$Y7 family). Figure **4** shows the curves and the MD data. The values of $R^2$ for the Hou *et al.* model of G$n$Y7 structures are 0.8994 and 0.4525 for the Young's and shear moduli, respectively. The low $R^2$ value for the shear modulus from Hou *et al.* G$n$Y7 model comes from the large difference between the MD data and the value from the model for $n = 1$. Although $R^2 < 0$, visually the Hou *et al.* [42] model is qualitatively fair for the Poisson's ratio of G$n$Y7 structures.

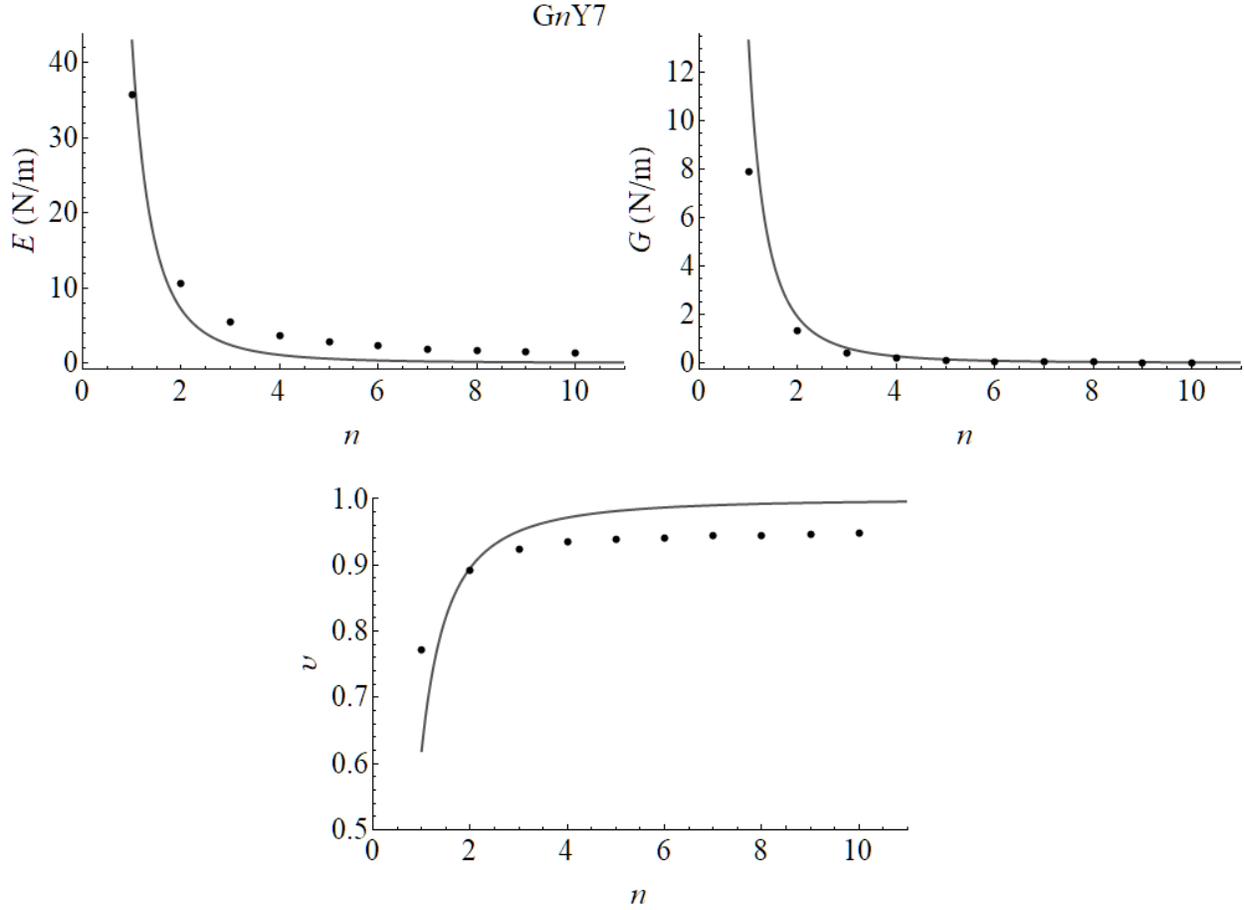

**Figure 4**: Young's modulus (top left), shear modulus (top right) and Poisson's ratio (bottom) of all "α-" or G$n$Y7 structures as function of $n$. MD data is shown in circles while full lines, from top left to bottom, represent the curves given by eq. (12), (14) and (13) of Hou *et al.* [42] model, respectively.

### 3.1.4 Discussion of the results for the elastic models for G$n$Y1, G$n$Y4 and G$n$Y7 structures

Based on the $R^2$ results and visual inspection of the results above, we can say that the models that best described the MD results for the dependence on $n$ of the elastic properties of some GFs are the Cranford *et al*. [39] and Hou et al. [42] models for the specific G$n$Y1 and G$n$Y7 families, respectively.

We have obtained data for all 10 members (with $1 \leq n \leq 10$) of all 7 originally proposed GFs ($1 \leq f \leq 7$). Instead of working on deducing specific mechanical models for specific GFs, we propose here a numerical approach to find out an *effective elastic model* that might be able to numerically describe, with the same formalism but possibly different coefficients, the dependence on $n$ of the elastic properties of the GY structures belonging to each GF. As the above Cranford *et al*. and Hou *et al*. models performed well within their corresponding GFs, they will be considered in the next subsection as fitting models to which we are going to try fitting all MD data and see which model provides the best numerical description of the dependence on $n$ of the elastic properties of all GYs. The successful *effective elastic model* is expected not only to fit very well the set of MD data for the Young's modulus that will be used to find out the values of fitting parameters, but also to describe with good precision the dependence on $n$ of the shear modulus and linear compressibility of all GYs, including the anisotropic ones. As a consequence of finding out such an *effective elastic model*, it will provide a physical interpretation for the elastic behavior of all GYs.

## 3.2 The *effective elastic model*

In this subsection, we will show how we found out the *effective elastic model* mentioned in the previous section. Although our *effective elastic model* cannot describe the Poisson's ratio of GYs, it will be shown to effectively describe the linear compressibility that is known to depend on the Poisson's ratio of the structure.

### 3.2.1 The fitting models

The two models that presented the best results for the dependence on *n* of the elastic properties of G*n*Y1 and G*n*Y7 structures, the Cranford *et al*. [39] and Hou et al. [42] models as shown in subsections **3.1.1** and **3.1.3**, respectively, will be considered here as basis for two fitting models called **M1** and **M2** models.

The model **M1** consists of a serial combination of *n* springs, and is based on the originally proposed model by Cranford *et al*. [39] for the dependence on *n* of the Young's modulus of the G*n*Y1 family. The model **M2** is based on stretching, hinging and bending deformations of the acetylene chains of GYs, and is taken from the model originally proposed by Hou *et al*. [42] for the elastic properties the G*n*Y7 family. The equations corresponding to models **M1** and **M2** are given below:

$$E_f^{M1}(n) = E_1^f a_1^f \left(a_1^f + \delta_f (n-1)\Delta a^f\right)^{-1}, \qquad (5)$$

$$E_f^{M2}(n) = \frac{4}{\sqrt{3}}\left(\gamma_f \frac{3}{L_1(n)} + \alpha_f \frac{L_2(n)*r_0^2}{6C_H} + \beta_f \frac{L_3(n)*r_0^2}{2C_B}\right)^{-1}, \qquad (6)$$

where $E_1^f$, $a_1^f$ and $\Delta a^f$ are the Young's modulus and lattice parameters for the G1Y*f* structures. In the original Cranford *et al*. model, $E_1^1$ and $a_1^1$ are the Young's modulus and lattice parameter of the G1Y1 structure, and $\Delta a^1$ = 2.66 Å [39]. The term $(a_1 + (n-1)\Delta a)$, without the superscript *f* and with $\delta_f = 1$ in eq. (5) is, in fact, the Cranford *et al*. proposed dependence on *n* of the lattice parameter of the G*n*Y1 structures. $E_1^f$, $a_1^f$ and $\Delta a^f$ for all G*n*Y*f*s obtained from our MD simulations will be used in eq. (5). Their values are shown in Tables **S1**, **S2.1** and **S2.2** of SM. Cranford *et al*. equation came from modeling the chain of *n* acetylene linkages in G*n*Y1s as a serial combination of *n* springs. The **M1** model of eq. (5) consists of a simple modification of the Cranford *et al.* model by including the parameter $\delta_f$ that is going to be adjusted based on MD data for the Young's modulus of the G$n$Y$f$ structures. $r_0$, $C_H$ and $C_B$ in eq. (6) are the length of the carbon-carbon bond in graphene, and the elastic constants related to bond angle rotations corresponding to hinging and bending of the acetylene chains, respectively. $r_0$ = 1.4 Å (from our MD simulations) and $C_H$ and $C_B$ values were taken from Ref. [42]. $L_1(n)$, $L_2(n)$ and $L_3(n)$ are functions of *n* and carbon-carbon bond distances in GYs that correspond to a given deformation: stretching, hinging and bending of the acetylene chains, respectively. They are reproduced in Section **S3** of the SM. The GF dependent $\gamma_f$, $\alpha_f$ and $\beta_f$ parameters are added to each term of the Hou *et al.* model to define the **M2** model of eq. (6). Each parameter multiplies one term corresponding to one kind of deformation in the model. $\gamma_f$, $\alpha_f$ and $\beta_f$ will be adjusted for each GF based on the MD data.

**3.2.2 Fitting the models to Young's modulus of all GYs**

The MD data for the Young's modulus of all GYs from all GFs are fitted by eqs. (5) and (6). The Young's modulus along the x and y directions of anisotropic G$n$Y$f$ structures will be fitted separately. Table **S4** of Section **S4** of SM shows the values of $\delta_f$, $\gamma_f$, $\alpha_f$ and $\beta_f$ for every G$n$Y$f$ family, including different values along x and y directions for the anisotropic GYs. $R^2$ values corresponding to these fittings are also shown for models **M1** and **M2** in Table **S4** of SM. Section **S4** of SM also describes how the fittings were obtained.

One would expect that by having more terms and more fitting parameters, **M2** would better succeed than **M1** in the description of the elastic properties of all GYs. In fact, in terms of the $R^2$ values of the fitting parameters (see Table **S4** of Section **S4** of SM for the numbers), the quality of the fittings **M1** and **M2** models to the MD data of the Young's modulus of all GYs are almost the same. However, as Figure **5** shows, while **M1** model visually fits quite well all the MD data for the Young's modulus of all GYs, including those along x and y directions for anisotropic structures, the **M2** model failed to fit the MD data for the Young's modulus along x (armchair) direction of the structures belonging to G$n$Y5 and G$n$Y6 families.

These results do not mean that **M2** is a bad fitting model. Some of the **M2** model $R^2$ values for some GFs are a little bit closer to 1 than the corresponding ones of the **M1** model. But it is surprising that the simplest physical model is able to fit quite well the Young's modulus of all GYs, including the anisotropic ones along x and y directions.

In spite of these good results for the **M1** model, before concluding that it really represents a good *effective elastic model* for the mechanisms behind the elastic properties of all GYs, we will perform a kind of *transferability* test of the model. Using the same fitted $\delta_f$ parameters, we are going to test the ability of **M1** model to fit the MD data for the shear modulus. That is presented in the next subsection.

### 3.2.3 Test of the M1 fitting model to shear modulus of all GYs

Here, we perform an important test for the ability of the **M1** model to describe the dependence on $n$ of the elastic properties of all GYs. $R^2$ values of the shear modulus version of the **M1** model that uses the same exact fitted $\delta_f$ parameters obtained with the MD data for the Young's modulus of all GYs, will be calculated using the MD data for their shear modulus. The expression that represents the **M1** model for the shear modulus of all GYs, that is similar to eq. (5), is given by:

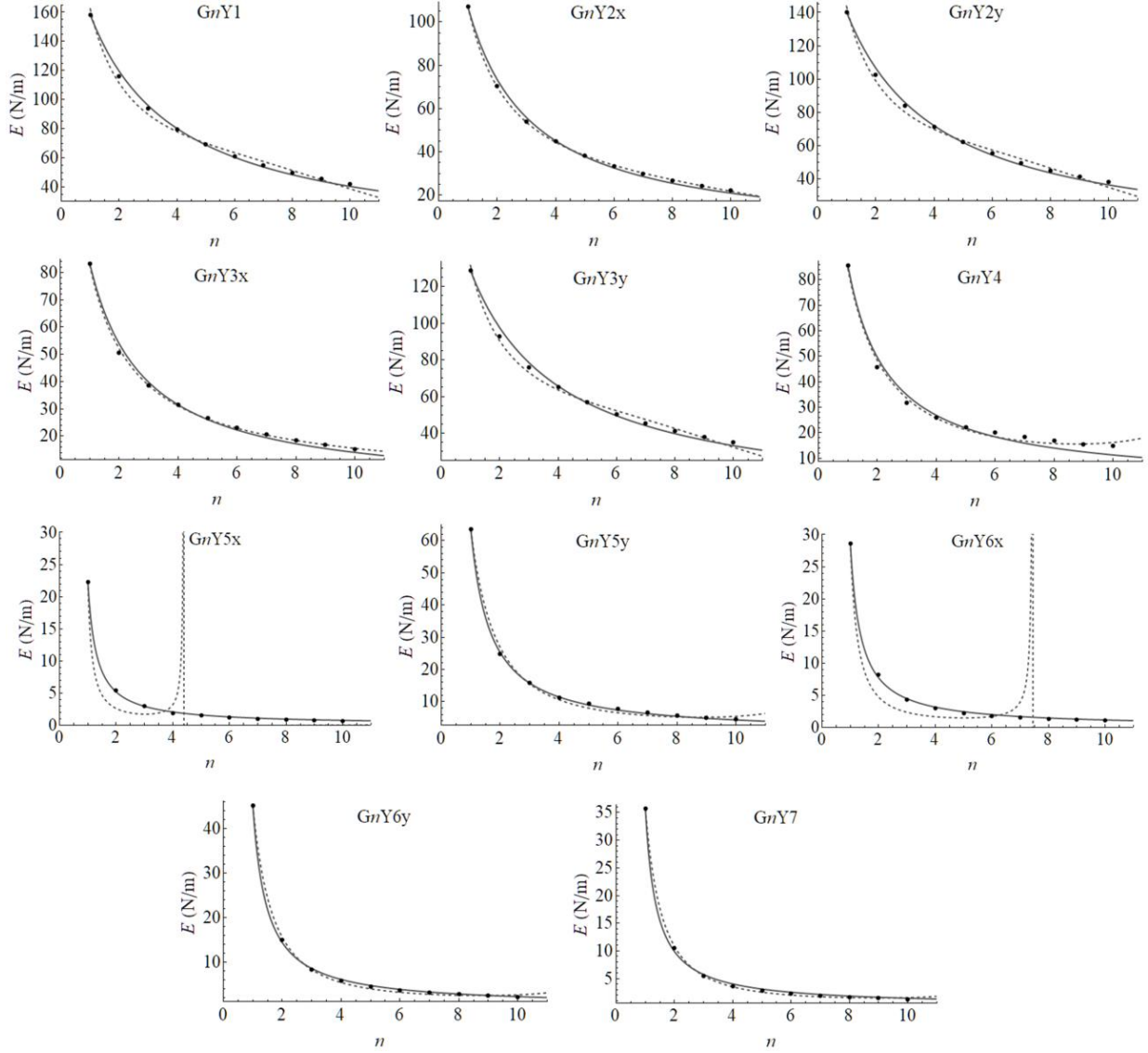

**Figure 5**: Young's modulus of all G$n$Y$f$ structures as function of $n$. In every plot, MD data is shown in circles and the full (dashed) line represents the eq. (5) (eq. (6)) for the fitted **M1** (**M2**) model. Plots of the Young's modulus along x and y directions are shown for anisotropic G$n$Y$f$ structures with appended "x" or "y" letter to the plot names.

$$G_f^{\mathbf{M1}}(n) = G_1^f a_1^f \left(a_1^f + \delta_f (n-1)\Delta a^f\right)^{-1} , \tag{7}$$

where $a_1^f$ and $\Delta a^f$ are the same as those used in eq. (5), and $G_1^f$ is the shear modulus of the corresponding G1Y$f$ structure.

Figure **6** shows the MD data and eq. (7) for the shear modulus of all G*n*Y*f* structures as well as the corresponding $R^2$ values. The smallest $R^2$ values are ~ 0.80 and 0.87 for the shear modulus of G*n*Y2 and G*n*Y4 families, but it is ≥ 0.94 for the G*n*Y3, G*n*Y6 and G*n*Y7 families and > 0.99 for the G*n*Y1 one.

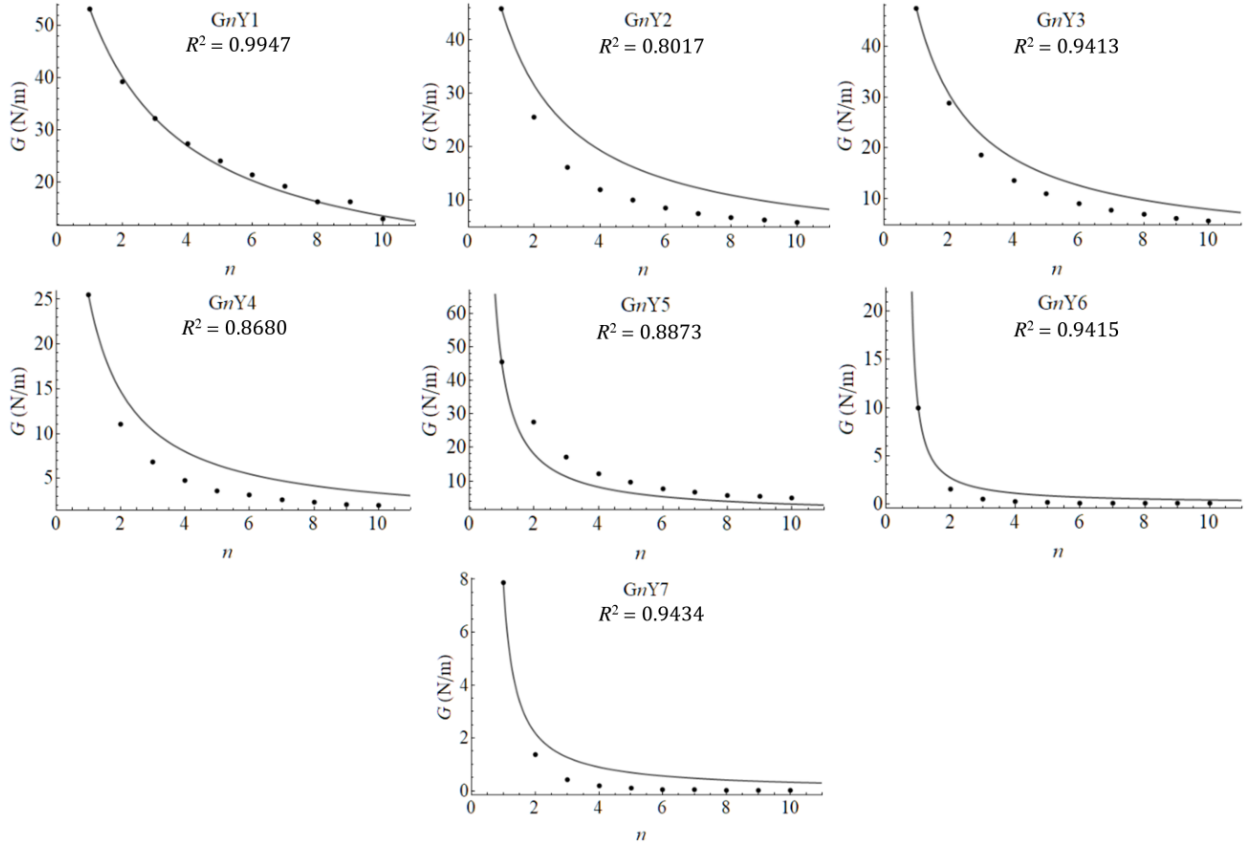

**Figure 6**: Shear modulus of all G*n*Y*f* structures as function of *n*. In every plot, MD data is shown in circles and the full line represents the eq. (7) with the same values of $\delta_f$ fitted for the Young's modulus of the corresponding G*n*Y*f*s. The value of the calculated $R^2$ is presented for each G*n*Y*f*.

Based on the above results, we conclude that the eq. (7), as well as the **M1** model, is good to describe the shear modulus of all GYs.

### 3.2.4 The Poisson's ratio of all GYs

The MD data for the Poisson's ratio of all GYs are shown in Figure **7**.

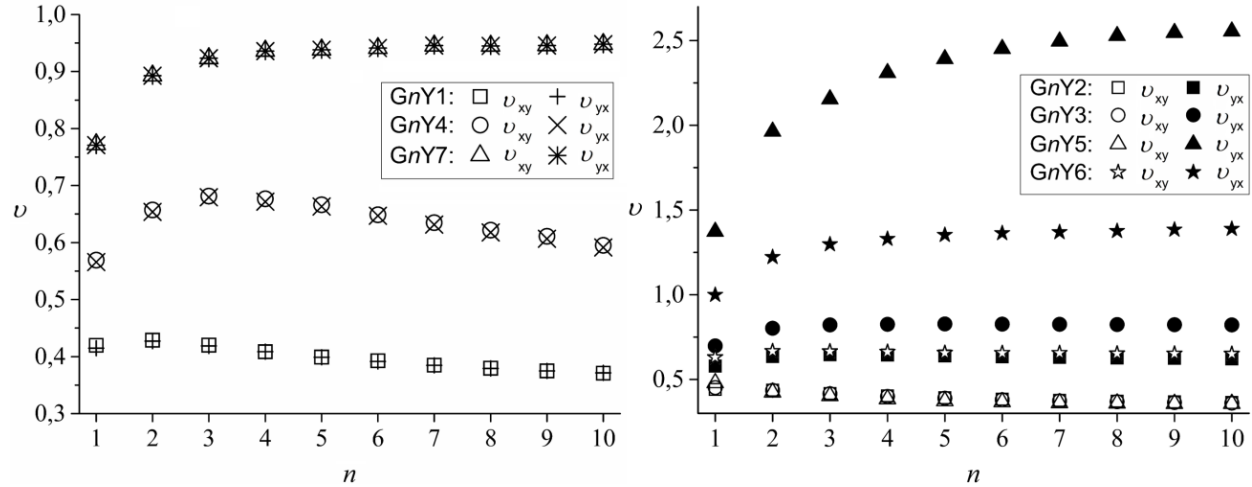

**Figure 7**: Poisson's ratio, $v$, of all symmetric (left panel) and asymmetric (right panel) G$n$Y$f$ structures. In order to verify the validity of the computational method, the results for $v_{xy}$ and $v_{yx}$ are shown for symmetric and asymmetry structures. See the legends for the data corresponding to each GF.

The Poisson's ratio, $v$, of all GYs are equal or larger than that of graphene (each value obtained from MD simulations is given in Table **S1** of SM). It means that the GY structures are relatively softer than graphene. As expected, the x and y values of $v$ for the symmetric GY structures (families G$n$Y1, G$n$Y4 and G$n$Y7) are the same (panel left of Figure **7**). Except for the G$n$Y4 family, all values of Poisson's ratio converges to a certain value with $n \to 10$. G$n$Y7 family presents the largest values of $v$ for symmetric GYs. Their Poisson's ratio converges with $n$ to a value a little bit smaller than 1. To our knowledge, the value of $v = 0.95$ for the G10Y7 structure is the largest one predicted for the Poisson's ratio of 2D symmetric crystals [51].

Amongst the asymmetric GYs, G$n$Y5 GF presents the largest values of $v_{yx}$ followed by the $v_{yx}$ values of the G$n$Y6 family (panel right of Figure 7). The smallest $v_{xy}$ are along the armchair direction of G$n$Y2, G$n$Y3 and G$n$Y5 GFs.

As shown by Grima *et al.* [60], the Poisson's ratio of a honeycomb structure subject only to stretching is negative. This auxetic property was, for example, taken into account in a proposed wine-rack-like structure of springs connected by hinges [61] to model the anisotropic Poisson's ratio of a complex system [62]. As shown in Figure 7, the Poisson's ratio all GYs are positive. Therefore, this property cannot be modeled only by a serial combination of springs. The Poisson's ratio requires a more specific modeling. Although this might weaken the proposal of the **M1** model as a good *effective elastic model* of all GYs, the analysis of the MD results for the linear compressibility surprisingly shows that the **M1** model is, in fact, good. It will be shown in the next subsection.

**3.2.5 The linear compressibility of all GYs**

Eqs. (2), (3) and (4) can be used to show that the linear compressibility depends on the Young's modulus, $E$, and Poisson's ratio, $v$, of the structure through $\beta_i = (E_i)^{-1} - v_{ji}(E_j)^{-1}$ [60], where $i$ and $j$ stand for x and y directions. Using the **M1** model given by eq. (5), the following expressions for the linear compressibility of the G$n$Y$f$s can be derived:

$$\beta^f(n) = \frac{1-v^f}{E_1^f} + \frac{(1-v^f)\delta a^f}{E_1^f a_1^f}(n-1) \, , f = 1, 4 \text{ and } 7, \tag{8a}$$

$$\beta_i^f(n) = \left(\frac{1}{E_{1i}^f} - \frac{v_{ji}^f}{E_{1j}^f}\right) + \left(\frac{\delta a_i^f}{E_{1i}^f a_{1i}^f} - \frac{v_{ji}^f \delta a_j^f}{E_{1j}^f a_{1j}^f}\right)(n-1) \, , f = 2, 3, 5 \text{ and } 6, \tag{8b}$$

where $\delta a^f \equiv \delta_f \Delta a^f$ (same $\delta_f$ and $\Delta a^f$ obtained from the previous fittings) and the choice of use of eqs. (8a) or (8b) depends only on the symmetry of the G$n$Y$f$. $i$ and $j$ in eq. (8b) represent x or y directions. Amongst the parameters and coefficients present in eqs. (8a) and (8b), the only one that implicitly depends on $n$ is the Poisson's ratio, $v^f$ and $v_{ji}^f$, respectively. As can be seen in Fig. **7**, the Poisson's ratio does not vary with $n$ by one order of magnitude or more as the Young's and shear moduli. Then, assuming for instance that the Poisson's ratio is roughly about constant, eqs. (8a) and (8b) predict a linear dependence on $n$ of the linear compressibility. Figure **8** shows the MD data for the linear compressibility of all GYs. Also, a linear fitting of the data corresponding to each GF is presented. The results remarkably show that the linear compressibility of all GYs are well described by a linear function of $n$. The $R^2$ values of these fittings are all close to 1. Table **1** shows the values of the fitted angular coefficients of the lines shown in Figure **8** together with the values of the coefficients that multiply $n$ in eqs. (8a) and (8b) considering the average values of the Poisson's ratio for each GF. Except for the linear compressibility along zigzag direction of the G$n$Y3 structures, the fitted angular coefficients agree with the coefficients from the **M1** model, including the negative angular coefficients of the linear compressibility of G$n$Y5 and G$n$Y6 structures along the zigzag or y direction. These results, together with those for Young's and shear moduli, indicate that the **M1** model is able to describe, with good precision, the elastic properties of all GYs.

Degabriele *et al*. [50] have predicted negative values for the linear compressibility of structures belonging to the G$n$Y5 family with $n$ from 1 to 5. Here, besides showing the linear trend, even for negative values, we also found out that G$n$Y6 family also present negative linear compressibility along the zigzag or y direction. In a recent review, Grima-Cornish *et al*. [63] have discussed the role of rotating rigid units and the negative linear compressibility in anisotropic structures possessing high (larger than 1) Poisson's ratio. Structures from G$n$Y5 and G$n$Y6 families satisfy these conditions. They are highly anisotropic, possess high Poisson's ratio along their zigzag directions and their sp2 carbon-carbon bonds might play the role of rotating rigid units.

**Table 1**: Values of the angular coefficients for the fitted lines, $\beta^f(n) = A_f + B_f n$, and the values of the coefficients that multiply $n$ in eqs. (8a) and (8b), using the corresponding average values of the Poisson's ratio, $\overline{v^f}$ for each G$n$Y$f$ family, and along x and y directions for asymmetric GFs. The $R^2$ values of each linear fit is also shown.

| $f$ and direction | $B_f$ | $\overline{v^f}$ | Eq. (8a) or (8b) coefficient | $R^2$ |
|---|---|---|---|---|
| 1 | 0.00125 | 0.398 | 0.00123 | 1.0 |
| 2x | 0.00257 | 0.395 | 0.00286 | 0.99971 |
| 2y | 0.000780 | 0.628 | 0.000553 | 0.99771 |
| 3x | 0.00386 | 0.394 | 0.00457 | 0.99943 |
| 3y | 0.000341 | 0.809 | -0.000105 | 0.95796 |
| 4 | 0.00248 | 0.636 | 0.00310 | 0.99998 |
| 5x | 0.1142 | 0.386 | 0.0951 | 0.99815 |
| 5y | -0.03824 | 2.276 | -0.03384 | 0.99935 |
| 6x | 0.03747 | 0.654 | 0.03334 | 0.99932 |
| 6y | -0.02024 | 1.308 | -0.01500 | 0.99936 |
| 7 | 0.00375 | 0.919 | 0.00595 | 0.99997 |

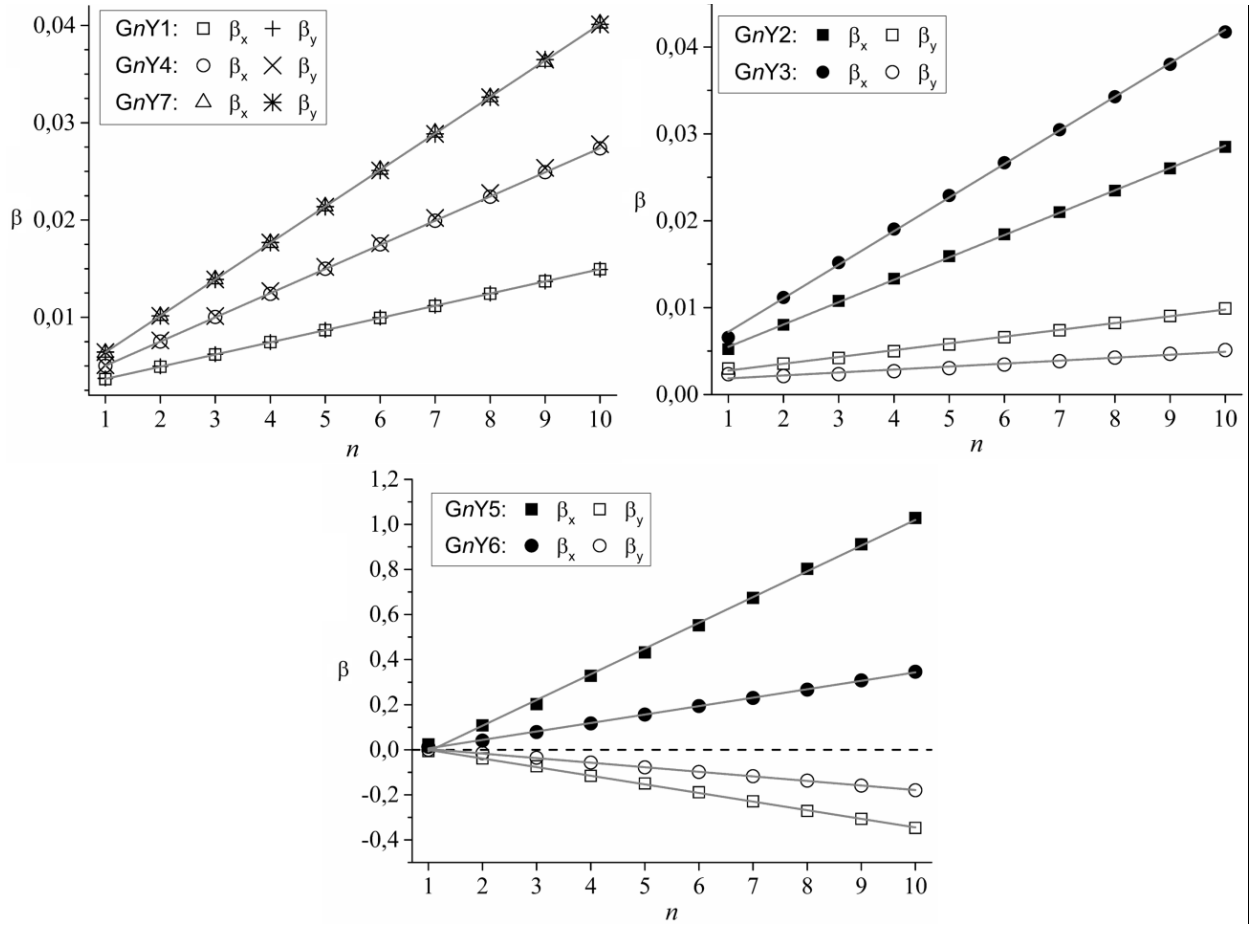

**Figure 8**: Linear compressibility, β, of all symmetric (top left panel) and asymmetric (top right and bottom panels) G$n$Y$f$ structures. In order to verify the validity of the computational method, the results for $β_x$ and $β_y$ are shown for symmetric and asymmetry structures. Dashed horizontal line in bottom panel represents $β = 0$ line and is shown to highlight the negative values of $β_y$ of G$n$Y5 and G$n$Y6 families.

## 4. Conclusions

Using MD simulations, we have calculated and presented the values of Young's modulus, shear modulus, Poisson's ratio and linear compressibility of 10 members of the 7 originally proposed families of GY structures. Each member of a GF is characterized by the number, $n$, of acetylene chains, with $1 \leq n \leq 10$ and each GF is represented by the parameter $f$. The set of MD data allowed us to verify existent models for the dependence on *n* of the elastic properties of some GFs. We have shown that the Young's modulus, shear modulus and linear compressibility of all GYs can be described, with good precision, by a simple *effective elastic model* based on a serial combination of *n* springs. The Poisson's ratio of the GYs cannot be modeled by this *effective elastic model* and requires more specific modeling. Further investigations about this property in GYs will be subject of future studies. Although not being able to describe the Poisson's ratio, it is remarkable that a simple model is capable to numerically describe the elastic properties of very different structural GYs, including anisotropy in some families. In the discussion of results, we also considered remarkable that this simple model, that cannot capture the dependence on *n* of the Poisson's ratio, is able to describe numerically well the dependence on *n* of the linear compressibility of different GY families, including the negative values found out along zigzag or y direction of G*n*Y5 and G*n*Y6 structures. In Cranford *et al*. [39] model, the springs are related to each acetylene chain covering a certain amount of surface. Here, eq. (5) cannot capture the same geometrical analysis made by Cranford *et al*. but one might think of G*n*Y*f*s being modeled by effective springs of different sizes and elastic constants amongst the families but acting similarly in the determination of the corresponding elastic properties. The MD data will be certainly useful to further test the development of new specific models for specific GY families. The present *effective elastic model* will be very useful to guide the design of applications of different GYs by easily predicting their mechanical behavior, especially for those GY structures for which there is not any specific model.

**Acknowledgements**

G. B. K. was supported by the Brazilian agency CNPq. A.F.F. is a fellow of the Brazilian Agency CNPq (#311587/2018-6) and acknowledges grant #2020/02044-9 from São Paulo Research Foundation (FAPESP). This research used the computing resources and assistance of the John David Rogers Computing Center (CCJDR) in the Institute of Physics "Gleb Wataghin", University of Campinas.

Supplementary Material for

# Effective acetylene length dependence of the elastic properties of different kinds of graphynes

*Guilherme B. Kanegae, Alexandre F. Fonseca* *

Applied Physics Department, Institute of Physics "Gleb Wataghin", University of Campinas - UNICAMP, 13083-859, Campinas, São Paulo, Brazil.

**Corresponding Author**

* Phone: +55 19 3521-5364. Email: Alexandre F. Fonseca – afonseca@ifi.unicamp.br

Here, the Supplemental Materials of the manuscript are presented and described.

**S1 Elastic constants of all GYs.** The values for $C_{11}$, $C_{12}$, $C_{22}$ and $C_{66}$ of all G$n$Y$f$s, with $1 \leq n \leq 10$ and $1 \leq f \leq 7$, as well as their corresponding values of Young's modulus, $E$, shear modulus, $G$, Poisson's ratio, $\nu$, and linear compressibility, $\beta$, are shown in the table below:

Table S1. Elastic constants, $C_{ij}$ (N/m), Young's, $E$, and shear, $G$, moduli (N/m), Poisson's ratio and linear compressibility (m/N) of all G$n$Y$f$ structures. If the structure is asymmetric, the elastic parameters along x (y) direction are shown outside (inside) parentheses.

| Structure | $C_{11}$ | $C_{12}$ | $C_{22}$ | $C_{66}$ | $E$ | $G$ | $\nu$ | $\beta$ |
|---|---|---|---|---|---|---|---|---|
| G1Y1 | 191.67 | 79.47 | 189.29 | 53.277 | 158.30 | 53.277 | 0.42 | 0.004 |
| G1Y2 | 143.79 | 83.22 | 188.45 | 45.933 | 107.04 (140.28) | 45.933 | 0.44 (0.58) | 0.005 (0.003) |
| G1Y3 | 121.69 | 84.85 | 187.98 | 47.336 | 83.39 (128.81) | 47.336 | 0.45 (0.70) | 0.007 (0.002) |
| G1Y4 | 126.51 | 71.59 | 125.77 | 25.452 | 85.77 | 25.452 | 0.57 | 0.005 |
| G1Y5 | 64.98 | 89.15 | 185.99 | 45.628 | 22.25 (63.68) | 45.628 | 0.48 (1.37) | 0.023 (-0.006) |
| G1Y6 | 77.29 | 77.22 | 122.41 | 10.001 | 28.58 (45.26) | 10.001 | 0.63 (1.00) | 0.013 (0.00002) |
| G1Y7 | 88.20 | 68.00 | 88.05 | 7.882 | 35.68 | 7.882 | 0.77 | 0.006 |
| G2Y1 | 141.82 | 60.58 | 141.20 | 39.267 | 115.83 | 39.267 | 0.43 | 0.005 |
| G2Y2 | 97.26 | 61.81 | 141.73 | 25.592 | 70.31 (102.45) | 25.592 | 0.44 (0.64) | 0.008 (0.004) |
| G2Y3 | 77.65 | 62.27 | 142.92 | 28.862 | 50.52 (92.98) | 28.862 | 0.44 (0.80) | 0.011 (0.002) |
| G2Y4 | 80.01 | 52.30 | 79.57 | 11.032 | 45.63 | 11.032 | 0.66 | 0.008 |
| G2Y5 | 31.93 | 62.69 | 147.84 | 27.674 | 5.35 (24.79) | 27.674 | 0.42 (1.96) | 0.108 (-0.039) |
| G2Y6 | 43.84 | 53.56 | 80.35 | 1.512 | 8.14 (14.92) | 1.512 | 0.67 (1.22) | 0.041 (-0.015) |
| G2Y7 | 52.03 | 46.45 | 52.03 | 1.359 | 10.55 | 1.359 | 0.89 | 0.010 |
| G3Y1 | 113.85 | 47.65 | 113.42 | 32.193 | 93.82 | 32.193 | 0.42 | 0.006 |
| G3Y2 | 74.13 | 47.91 | 114.92 | 16.164 | 54.15 (83.95) | 16.164 | 0.42 (0.65) | 0.011 (0.004) |
| G3Y3 | 58.44 | 47.99 | 115.52 | 18.635 | 38.50 (76.11) | 18.635 | 0.42 (0.82) | 0.015 (0.002) |
| G3Y4 | 59.11 | 40.16 | 58.93 | 6.825 | 31.74 | 6.825 | 0.68 | 0.010 |
| G3Y5 | 22.21 | 47.85 | 118.83 | 17.127 | 2.94 (15.73) | 17.127 | 0.40 (2.15) | 0.203 (-0.073) |
| G3Y6 | 30.96 | 40.17 | 60.45 | 0.461 | 4.27 (8.33) | 0.461 | 0.66 (1.30) | 0.079 (-0.036) |

Table S1. Continuation.

| Structure | $C_{11}$ | $C_{12}$ | $C_{22}$ | $C_{66}$ | $E$ | $G$ | $\nu$ | $\beta$ |
|---|---|---|---|---|---|---|---|---|
| G3Y7 | 37.39 | 34.52 | 37.40 | 0.421 | 5.52 | 0.421 | 0.92 | 0.014 |
| G4Y1 | 95.29 | 38.94 | 95.26 | 27.362 | 79.37 | 27.362 | 0.41 | 0.007 |
| G4Y2 | 60.43 | 38.87 | 96.41 | 11.978 | 44.76 (71.41) | 11.978 | 0.40 (0.64) | 0.013 (0.005) |
| G4Y3 | 47.03 | 38.83 | 97.11 | 13.651 | 31.51 (65.06) | 13.651 | 0.40 (0.83) | 0.019 (0.003) |
| G4Y4 | 47.75 | 32.07 | 47.40 | 4.780 | 26.04 | 4.780 | 0.68 | 0.012 |
| G4Y5 | 16.66 | 38.47 | 100.11 | 12.121 | 1.88 (11.28) | 12.121 | 0.38 (2.31) | 0.328 (-0.116) |
| G4Y6 | 24.11 | 32.03 | 48.40 | 0.195 | 2.90 (5.82) | 0.195 | 0.66 (1.33) | 0.117 (-0.056) |
| G4Y7 | 29.23 | 27.36 | 29.23 | 0.179 | 3.62 | 0.179 | 0.94 | 0.018 |
| G5Y1 | 82.22 | 32.80 | 82.14 | 24.076 | 69.12 | 24.076 | 0.40 | 0.009 |
| G5Y2 | 51.01 | 32.61 | 83.29 | 9.969 | 38.24 (62.43) | 9.969 | 0.39 (0.64) | 0.016 (0.006) |
| G5Y3 | 39.33 | 32.54 | 83.72 | 11.060 | 26.68 (56.80) | 11.060 | 0.39 (0.83) | 0.023 (0.003) |
| G5Y4 | 39.89 | 26.47 | 39.71 | 3.553 | 22.25 | 3.553 | 0.67 | 0.015 |
| G5Y5 | 13.46 | 32.18 | 86.24 | 9.632 | 1.45 (9.29) | 9.632 | 0.37 (2.39) | 0.432 (-0.150) |
| G5Y6 | 19.67 | 26.60 | 40.51 | 0.102 | 2.20 (4.53) | 0.102 | 0.66 (1.35) | 0.156 (-0.078) |
| G5Y7 | 24.12 | 22.63 | 24.12 | 0.093 | 2.88 | 0.093 | 0.94 | 0.021 |
| G6Y1 | 72.23 | 28.28 | 71.98 | 21.522 | 61.12 | 21.522 | 0.39 | 0.010 |
| G6Y2 | 44.22 | 28.06 | 73.20 | 8.488 | 33.46 (55.39) | 8.488 | 0.38 (0.63) | 0.018 (0.007) |
| G6Y3 | 33.88 | 27.98 | 73.61 | 9.154 | 23.24 (50.50) | 9.154 | 0.38 (0.83) | 0.027 (0.003) |
| G6Y4 | 34.66 | 22.39 | 34.48 | 3.125 | 20.12 | 3.125 | 0.65 | 0.017 |
| G6Y5 | 11.29 | 27.67 | 75.49 | 7.655 | 1.15 (7.68) | 7.655 | 0.37 (2.45) | 0.552 (-0.189) |
| G6Y6 | 16.69 | 22.75 | 34.72 | 0.087 | 1.78 (3.70) | 0.087 | 0.66 (1.36) | 0.194 (-0.098) |
| G6Y7 | 20.49 | 19.29 | 20.50 | 0.055 | 2.34 | 0.055 | 0.94 | 0.025 |
| G7Y1 | 64.55 | 24.83 | 64.49 | 19.339 | 54.99 | 19.339 | 0.38 | 0.011 |
| G7Y2 | 38.98 | 24.61 | 65.36 | 7.495 | 29.71 (49.83) | 7.495 | 0.38 (0.63) | 0.021 (0.007) |
| G7Y3 | 29.72 | 24.53 | 65.71 | 7.861 | 20.56 (45.47) | 7.861 | 0.37 (0.83) | 0.030 (0.004) |
| G7Y4 | 30.59 | 19.32 | 30.43 | 2.576 | 18.32 | 2.576 | 0.64 | 0.020 |
| G7Y5 | 9.73 | 24.27 | 67.06 | 6.739 | 0.95 (6.53) | 6.739 | 0.36 (2.49) | 0.674 (-0.229) |
| G7Y6 | 14.52 | 19.87 | 30.33 | 0.039 | 1.50 (3.14) | 0.039 | 0.66 (1.37) | 0.230 (-0.117) |

Table S1. Continuation.

| Structure | $C_{11}$ | $C_{12}$ | $C_{22}$ | $C_{66}$ | $E$ | $G$ | $\nu$ | $\beta$ |
|---|---|---|---|---|---|---|---|---|
| G7Y7 | 17.77 | 16.81 | 17.78 | 0.035 | 1.89 | 0.035 | 0.95 | 0.029 |
| G8Y1 | 58.29 | 22.11 | 58.24 | 16.250 | 49.89 | 16.250 | 0.38 | 0.012 |
| G8Y2 | 34.92 | 21.90 | 58.95 | 6.767 | 26.78 (45.21) | 6.767 | 0.37 (0.63) | 0.023 (0.008) |
| G8Y3 | 26.48 | 21.83 | 59.28 | 6.933 | 18.44 (41.29) | 6.933 | 0.37 (0.82) | 0.034 (0.004) |
| G8Y4 | 27.43 | 16.94 | 27.25 | 2.307 | 16.90 | 2.307 | 0.62 | 0.022 |
| G8Y5 | 8.55 | 21.61 | 60.26 | 5.740 | 0.80 (5.63) | 5.740 | 0.36 (2.53) | 0.803 (-0.271) |
| G8Y6 | 12.82 | 17.64 | 27.00 | 0.027 | 1.30 (2.74) | 0.027 | 0.65 (1.38) | 0.267 (-0.137) |
| G8Y7 | 15.75 | 14.89 | 15.75 | 0.024 | 1.69 | 0.024 | 0.94 | 0.033 |
| G9Y1 | 53.12 | 19.92 | 53.14 | 16.270 | 45.65 | 16.270 | 0.37 | 0.014 |
| G9Y2 | 31.58 | 19.73 | 53.80 | 6.220 | 24.34 (41.48) | 6.220 | 0.37 (0.62) | 0.026 (0.009) |
| G9Y3 | 23.89 | 19.66 | 54.06 | 6.245 | 16.74 (37.88) | 6.245 | 0.36 (0.82) | 0.038 (0.005) |
| G9Y4 | 24.80 | 15.05 | 24.63 | 2.103 | 15.61 | 2.103 | 0.61 | 0.025 |
| G9Y5 | 7.66 | 19.49 | 54.66 | 5.373 | 0.71 (5.04) | 5.373 | 0.36 (2.55) | 0.912 (-0.307) |
| G9Y6 | 11.46 | 15.85 | 24.34 | 0.020 | 1.13 (2.41) | 0.020 | 0.65 (1.38) | 0.307 (-0.159) |
| G9Y7 | 14.13 | 13.36 | 14.13 | 0.017 | 1.49 | 0.017 | 0.95 | 0.036 |
| G10Y1 | 48.78 | 18.12 | 48.85 | 12.985 | 42.06 | 12.985 | 0.37 | 0.015 |
| G10Y2 | 28.85 | 17.94 | 49.35 | 5.765 | 22.33 (38.19) | 5.765 | 0.36 (0.62) | 0.028 (0.010) |
| G10Y3 | 21.77 | 17.88 | 49.57 | 5.701 | 15.32 (34.89) | 5.701 | 0.36 (0.82) | 0.042 (0.005) |
| G10Y4 | 22.82 | 13.50 | 22.68 | 1.972 | 14.79 | 1.972 | 0.60 | 0.027 |
| G10Y5 | 6.95 | 17.76 | 49.85 | 4.977 | 0.63 (4.49) | 4.977 | 0.36 (2.55) | 1.029 (-0.346) |
| G10Y6 | 10.37 | 14.39 | 22.13 | 0.015 | 1.01 (2.16) | 0.015 | 0.65 (1.39) | 0.346 (-0.180) |
| G10Y7 | 12.78 | 12.12 | 12.78 | 0.013 | 1.29 | 0.013 | 0.95 | 0.040 |

**S2. Lattice parameters of all GY structures.** Lattice parameters of all symmetric and asymmetric graphyne (GY) structures were obtained from the MD simulations. From all data, we fitted eq. (S2.1) for every set of lattice values of each structure. In order to avoid confusion, in the next equations, subscript indices $n$ are related to the number of acetylene chains of the structure, ($1 \leq n \leq 10$) and superscript indices $f$ are related to the family number of the structure, ($1 \leq f \leq 7$). The proposed dependence on $n$ of the lattice parameter of the GY structures is given by:

$$a_n^f = a_1^f + (n-1)\Delta a^f \, , \tag{S2.1}$$

Tables **S2.1** and **S2.2** show the results for $a_1^f$ and $\Delta a^f$.

**Table S2.1** Values of $a_1^f$[Å] ($1 \leq f \leq 7$) obtained from the MD simulation and used to the fitting of $\Delta a^f$ of all G$n$Y$f$ structures according to the eq. (S2.1). For asymmetric GYs, the values along $x$ ($y$) direction are shown outside (inside) parentheses.

| $a_1^1$ | $a_1^2$ | $a_1^3$ | $a_1^4$ | $a_1^5$ | $a_1^6$ | $a_1^7$ |
|---|---|---|---|---|---|---|
| 6.90 | 6.90 (9.53) | 6.90 (26.17) | 9.61 | 6.89 (7.11) | 9.60 (7.11) | 12.31 |

**Table S2.2** Values of $\Delta a^f$[Å] ($1 \leq f \leq 7$) obtained from fitting the data for the lattice parameters of all G$n$Y$f$ structures to the eq. (S2.1). For asymmetric GYs, the values along $x$ ($y$) direction are shown outside (inside) parentheses. All fitting calculations provided $R^2 \geq 0.999998$.

| $\Delta a^1$ | $\Delta a^2$ | $\Delta a^3$ | $\Delta a^4$ | $\Delta a^5$ | $\Delta a^6$ | $\Delta a^7$ |
|---|---|---|---|---|---|---|
| 2.66 | 2.66 (4.61) | 2.66 (13.84) | 5.33 | 2.65 (4.62) | 5.32 (4.62) | 7.99 |

The quality of the fitting given by the values of $R^2$ shows that the dependence of the lattice parameter with $n$ is well described by the linear relationship given by eq. (S2.1). The value of $\Delta a^1 = 2.66$ Å is the same as the one obtained by Cranford *et al.* [SM1] using another molecular dynamics (MD) potential called ReaxFF [SM2].

**S3. The functions $L_1(n)$, $L_2(n)$ and $L_3(n)$.** Eq. (6) of the main text for the **M2** model for the dependence on $n$ of the Young's modulus of the G$n$Y7s, is written in terms of three functions of $n$: $L_1(n)$, $L_2(n)$ and $L_3(n)$. According to Hou *et al.* [SM3] (or Ref. [40] of the main text), they are given by:

$$L_1(n) = \frac{559}{1.85n + 1}, \qquad (S3.1)$$

$$L_2(n) = [q_b + n(q_b + q_c)]^2, \qquad (S3.2)$$

$$L_3(n) = (q_b + q_c)\left[\frac{n(n+1)(4n-1)}{12}(q_b + q_c) - \frac{2n+1-(-1)^n}{8}(q_b - q_c)\right], (S3.3)$$

where $q_b \equiv b/r_0$ and $q_c \equiv c/r_0$ are the ratios of the lengths of sp and sp3 carbon-carbon bonds by the carbon-carbon bond distance in graphene ($r_0 = 1.4$ Å). $q_b = 1$ $q_c = 0.85$ and the numerical parameters in the eq. (S3.1) are taken from Ref. [SM3].

**S4. Information about the fitting of the Young's modulus of the GYs by M1 and M2 models.** We have used the following fitting tools and commands from the Mathematica Wolfram software to obtain the fitting parameters of models **M1** and **M2**:

```
Clear[nlmg,nlmg2,deltaf,gamaf,alfaf,betaf];
(* nlmg and nlmg2 will retain the results for the fittings of eqs. (5) and
(6) of them main text, respectively. Below their definition in Mathematica
software. *)
nlmg = NonlinearModelFit [compEx1,E1*a1/(a1+deltaf*(n-1)*Deltaa),deltaf,n];

nlmg2 = NonlinearModelFit[compEx1,(4/Sqrt[3])/(3*gamaf/L1[n]+
alfaf*L2[n]*(r0)^2/(6*CH) + betaf*Re[L3[n]]*r0*r0/(2*CB) ),
{gamaf,alfaf,betaf},n];

(* deltaf = δ_f, gamaf = γ_f, alfaf = α_f and  betaf = β_f are the fitting
parameters given in eqs. (5) and (6) of them main text, and L1[n], L2[n] and
L3[n] are given by eqs. (S3.1), (S3.2) and (S3.3) at Section 3 of this SM. *)
```

Table **S4** shows the values of fitted parameters of eqs. (5) and (6) of the main text for the MD data of Young's modulus of all GY structures.

**Table S4**: $\delta_f$, $\gamma_f$, $\alpha_f$ and $\beta_f$ parameters for each G$n$Y$f$ family from the fitting of eqs. (5) and (6) of the main text by the MD Young's modulus data of all G$n$Y$f$ structures. Also, the $R^2$ values from the Mathematica calculations of the fitted parameters are given for the M1 and M2 models.

| $f$ and direction | $\delta_f$ | $R^2$ for **M1** | $\alpha_f$ | $\beta_f$ | $\gamma_f$ | $R^2$ for **M2** |
|---|---|---|---|---|---|---|
| 1 | 0.8399 | 0.999594 | − 0.1708 | 0.003200 | 1.1347 | 0.998886 |
| 2x | 1.1855 | 0.999216 | − 0.1509 | 0.002551 | 1.5873 | 0.999952 |
| 2y | 0.6498 | 0.999425 | − 0.1983 | 0.003700 | 1.2877 | 0.998926 |
| 3x | 1.4210 | 0.998629 | − 0.0904 | 0.000891 | 1.9343 | 0.999785 |
| 3y | 0.6043 | 0.999189 | − 0.2123 | 0.003913 | 1.4020 | 0.999148 |
| 4 | 1.3185 | 0.995111 | 0.2073 | − 0.007112 | 1.5445 | 0.998311 |
| 5x | 8.6385 | 0.999746 | 88.207 | − 4.8409 | − 94.794 | 0.956083 |
| 5y | 2.3455 | 0.999894 | 1.7257 | − 0.03971 | 0.3282 | 0.998234 |
| 6x | 4.9192 | 0.999541 | 29.179 | − 1.0415 | − 29.055 | 0.953445 |
| 6y | 3.3049 | 0.999797 | 4.1854 | − 0.09246 | − 1.65278 | 0.999360 |
| 7 | 4.0283 | 0.999372 | 6.9137 | − 0.1474 | − 4.0229 | 0.999723 |